\documentclass[a4paper]{jpconf}
\usepackage[dvipdfmx]{graphicx}
\usepackage{amsmath}

\newcommand{\hw}{\ensuremath{\hbar\omega}}
\newcommand{\TN}{\ensuremath{T_{\rm N}}}
\newcommand{\m}{Mn$_3$Pt}
\newcommand{\Tt}{\ensuremath{T_{\rm t}}}
\newcommand{\ibbs}[1]{\ensuremath{\boldsymbol{#1}}}
\newcommand{\Ei}{\ensuremath{E_{\rm i}}}

\begin{document}
\title{Spin wave dispersion just above the magnetic order-order transition in the metallic antiferromagnet {\m}}

\author{Soshi Ibuka$^1$, Tetsuya Yokoo$^1$, Shinichi Itoh$^1$, Kazuya Kamazawa$^2$, Mitsutaka Nakamura$^3$ and Motoharu Imai$^4$}

\address{$^1$ Institute of Materials Structure Science, High Energy Accelerator Research Organization, Tokai, Ibaraki 319-1106, Japan}
\address{$^2$ Center for Neutron Science and Technology, Comprehensive Research Organization for Science and Society, Tokai, Ibaraki 319-1106, Japan}
\address{$^3$ Materials and Life Science Division, J-PARC Center, Japan Atomic Energy Agency, Tokai, Ibaraki 319-1195, Japan}
\address{$^4$ National Institute for Materials Science, Tsukuba, Ibaraki 305-0047, Japan}

\ead{ibuka@post.j-parc.jp}

\begin{abstract}
Spin wave dispersion in the metallic antiferromagnet {\m} was investigated just above the order-order transition temperature by using the inelastic neutron scattering technique. The spin wave dispersion at $T = 400$~K along [100], [110] and [111] directions was isotropic within the measurement accuracy. The dispersion was described by $({\hw})^2 = c^2q^2 + \Delta^2$ with $c = 190$~meV{\AA} and $\Delta = 3.3$~meV. Compared with the dispersion at $T = 419$~K previously reported, the result demonstrates a large reduction of the stiffness constant $c$ with increasing temperature. This is similar to that observed in the metallic antiferromagnet FePt$_3$, and is an indication of the itinerancy of the magnetic moments.
\end{abstract}

\section{Introduction}
The recent discovery of iron-based superconductors~\cite{KamiharaY2008} has renewed interest in a basic understanding of the dynamical magnetic properties of metallic antiferromagnets. 
In metallic antiferromagnets with 4$d$ or 5$d$ electrons like FePt$_3$~\cite{KohgiM1980}, the spin dynamics are expected to show the intermediate character between itinerant electron magnets, such as Cr and $\gamma$-FeMn, and localized spin magnets, such as Heusler alloys, because of the high polarizability of 4$d$ and 5$d$ electrons~\cite{IshikawaY1978}. The understanding of the spin dynamics in these magnets is still incomplete. Thus, the aim of this study is to examine the itinerancy and the locality of the magnetic moments in one of these alloys, \m.

\begin{figure}
\begin{center}
\hspace{4pc}
\includegraphics[width=0.23\hsize]{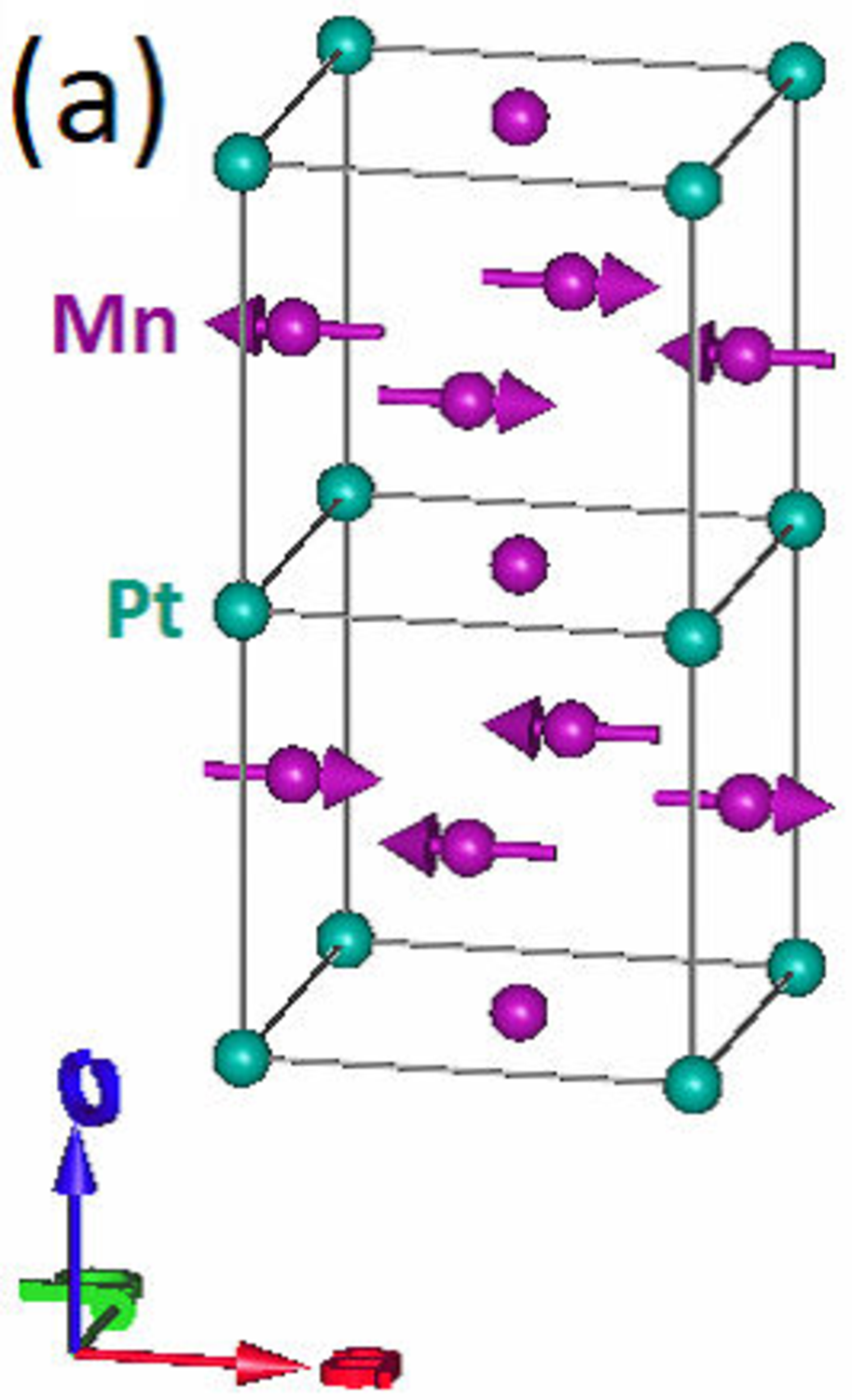}
\includegraphics[width=0.6\hsize]{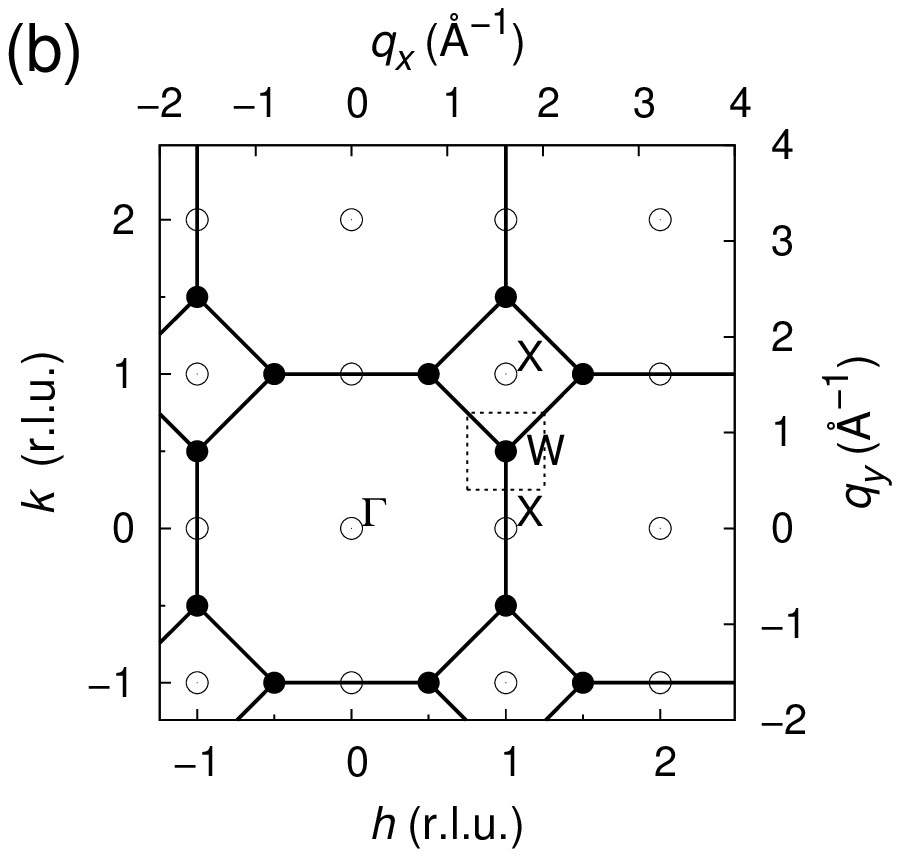}
\end{center}
\caption{\label{fig1}(a) Magnetic structure in the F-phase of {\m}. Pt ions do not have a magnetic moment. Two-third of Mn moments ordered with $\ibbs{Q}_{\rm F} = (1, 0, 0.5)$. (b) Brillouin zone of {\m} in $hk$0 plane. The open and close circles stand for the nuclear and magnetic elastic peak positions, respectively. The magnetic Brillouin zone was shown by the dotted line around a W point.}
\end{figure}
Mn$_3$Pt is a Cu$_3$Au-type metallic antiferromagnet with the transition temperature $\TN = 475$~K~\cite{KrenE1967, KrenE1968a}. It shows magnetic order-order transition at $\Tt = 400$~K. In the high-temperature ordered phase, which is known as F-phase, two third of Mn moments are ordered collinearly and one third of Mn moments are not ordered. The magnetic structure was shown in Fig.~\ref{fig1}(a). 
A slight tetragonal distortion $c/a = 0.9985$ occurs at \TN~\cite{IkedaT2003}, reflecting the magnetic symmetry of the partial disordered state. 
In the low temperature phase, which is called D-phase, Mn moments form a triangular spin structure with the moments lying in the 111 plane. The F- to D-phase transition leads to dramatic decreases in Mn moment from 3.4 to 2.2$\mu_{\rm B}$/Mn and in lattice volume by 2.25\%~\cite{KrenE1967, YasuiH1992}. 

The stability of the magnetic structures has been discussed with localized models. Kr\'{e}n et al have explained~\cite{KrenE1968a} the magnetic transition within the molecular-field approximation with the nearest-neighbor $J_1$ and the next-nearest-neighbor $J_2$ Mn-Mn exchange interactions, and showed that the exchange interactions are sensitive to the interatomic distances by studying the composition dependence of the magnetic structure. 
Ricodeau has introduced~\cite{RicodeauJA1974} the volume independent nearest-neighbor interaction $J$, and pointed out the importance of the taking into account of the entropy of the disordered Mn moments for stabilizing the F-phase. 
On the other hand, the spin wave in the D-phase shows itinerant characters experimentally. The spin wave stiffness is as much as 240~meV{\AA} at $T = 20$~K~\cite{TomiyoshiS1990}. The steep spin dispersion is characteristic of the itinerant electron magnet. In addition, large reduction of the stiffness constant with increasing temperature, similar to that in FePt$_3$, was observed in the D-phase~\cite{TomiyoshiS1990, YamaguchiY1992}.
In the F-phase, the stiffness constant is reduced to 130~meV{\AA} at 419~K, and 120~meV{\AA} at 459~K~\cite{TomiyoshiS1992, YamaguchiY1992}. 
With further increasing temperature, two-dimensional short-range correlations are developed~\cite{IkedaT2003, IkedaT2004, TomiyasuK2012}.
The itinerant character of the spin wave is seemingly weaker in the F-phase than in the D-phase. An investigation of the variation of the spin dynamics for a different temperature in the F-phase would give useful information to understand the spin dynamics of {\m}. Therefore, in this study, we performed inelastic neutron scattering experiments at $T = 400$~K just above \Tt.

\section{Experiments}
A single crystal of {\m} was synthesized by the Bridgman method. Mn chips and Pt wire were weighed with a molar ratio of Mn:Pt = 3.15:1. The purity was 99.99\% for Mn and 99.99\% for Pt. 
First, they were melted with using arc furnace under an argon gas atmosphere. The resulting alloy was placed in an alumina crucible, sealed in an electrically-fused quartz tube under an argon gas atmosphere, set in a vertical Bridgman furnace, and heated up to 1513~K. Next, it was slowly cooled down to 1413~K in 6 hours by moving it down to the bottom of the furnace. The temperature gradient of the furnace was 12.5~K/cm. Then, the sample was annealed at 1173~K for 90~h. The mass of the grown single crystal was 20~g. The chemical composition was determined by energy-dispersive X-ray spectroscopy with a scanning electron microscope (TM3030Plus and Quantax70, Hitachi High-Technologies). The chemical composition was Mn:Pt = 75.5(22):24.5(9). No impurity phase was found. The crystal orientation was determined by the x-ray Laue method with using YXLON MG452 (450~kV, 5~mA) and was aligned in $hk$0 zone. The Brillouin zone in $hk$0 plane was shown in Fig.~\ref{fig1}(b). Inelastic neutron scattering experiments were performed by using the chopper spectrometer 4SEASONS~\cite{KajimotoR2011, NakamuraM2009, InamuraY2013} installed at the Materials and Life Science Experimental Facility (MLF) in J-PARC, Japan. DAVE/MSlice~\cite{dave} was used for analyzing the data. Incident neutron energies were set to ${\Ei} = 10.2$, 14.8, 23.5, 42.9 and 102 meV with the energy resolution at the elastic position being 0.45, 0.7, 1.2, 2.5 and 8 meV, respectively. The crystal was sealed in an aluminum can under a $^4$He gas atmosphere and then set in a closed cycle $^4$He cryostat. Measurements were performed at $T = 400$~K. We confirmed that magnetic elastic peaks emerged at the W points, indicating that at least part of the sample was in F-phase.

\section{Results}
\begin{figure}
\begin{center}
\includegraphics[width=0.45\hsize]{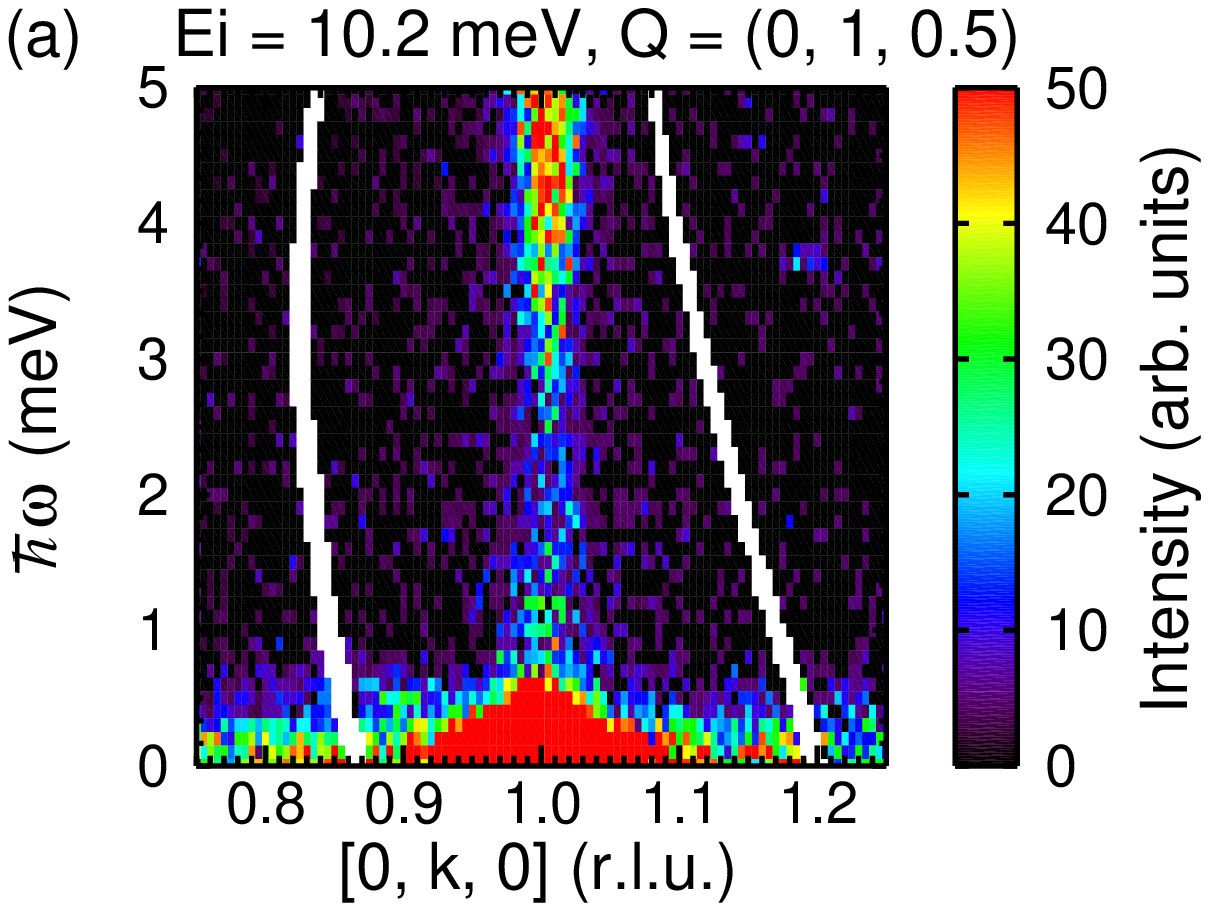}
\hspace{1pc}
\includegraphics[width=0.48\hsize]{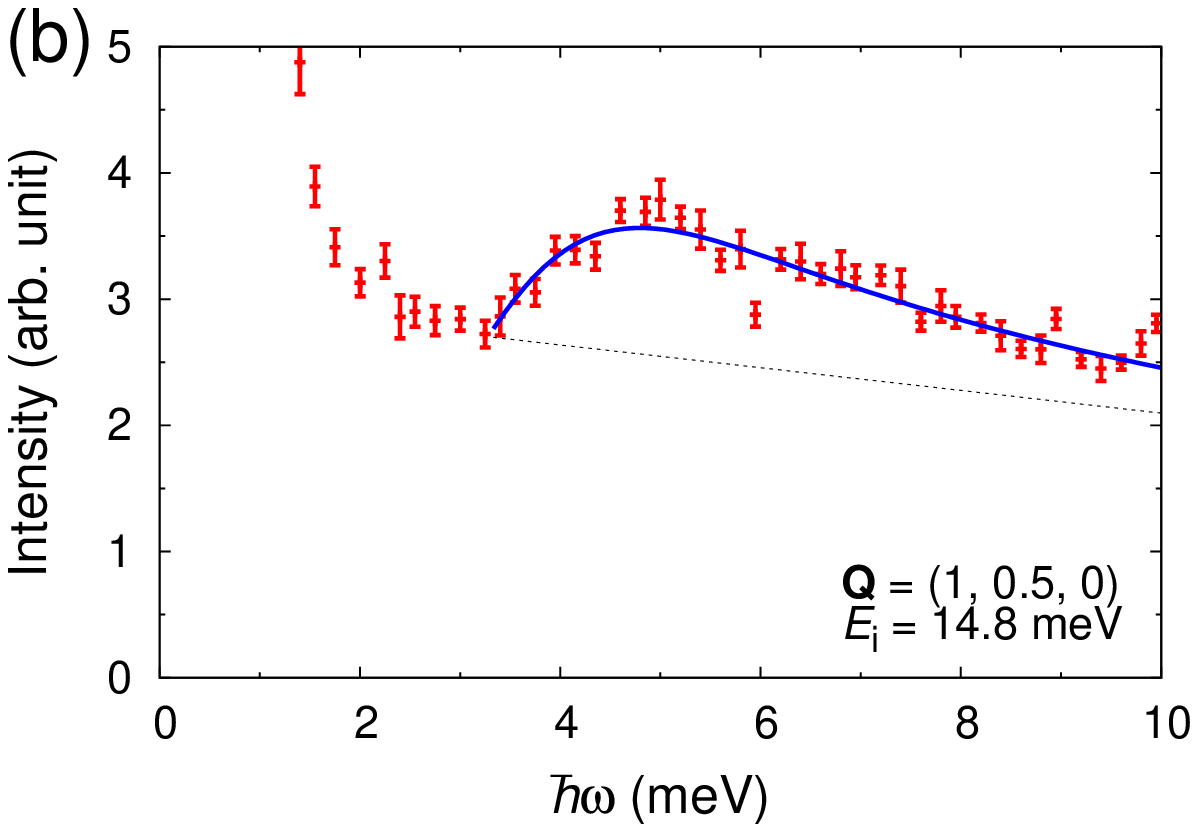}
\end{center}
\caption{\label{fig2}(a) Intensity map with the excitation energy {\hw} and $k$ around $\ibbs{Q} = (0, 1, 0.5)$ with ${\Ei} = 10.2$~meV. (b) {\hw} dependence of the scattering intensity at $\ibbs{Q} = (1, 0.5, 0)$  with ${\Ei} = 14.8$~meV. The solid line denotes the fit with a Lorentz function. The linear dotted line stands for the background.}
\end{figure}
First, an energy gap of the spin wave ${\Delta}$ was investigated by using low-energy incident neutrons. Figure~\ref{fig2}(a) shows the neutron scattering intensity map with the excitation energy {\hw} and $k$ around a W point. A spin excitation existed at $k$ = 1. Figure~\ref{fig2}(b) presents {\hw} dependence of the scattering intensity at $\ibbs{Q} = (1, 0.5, 0)$. ${\Delta}$ was estimated to be 3.3(1)~meV. 

\begin{figure}
\begin{center}
\includegraphics[width=0.45\hsize]{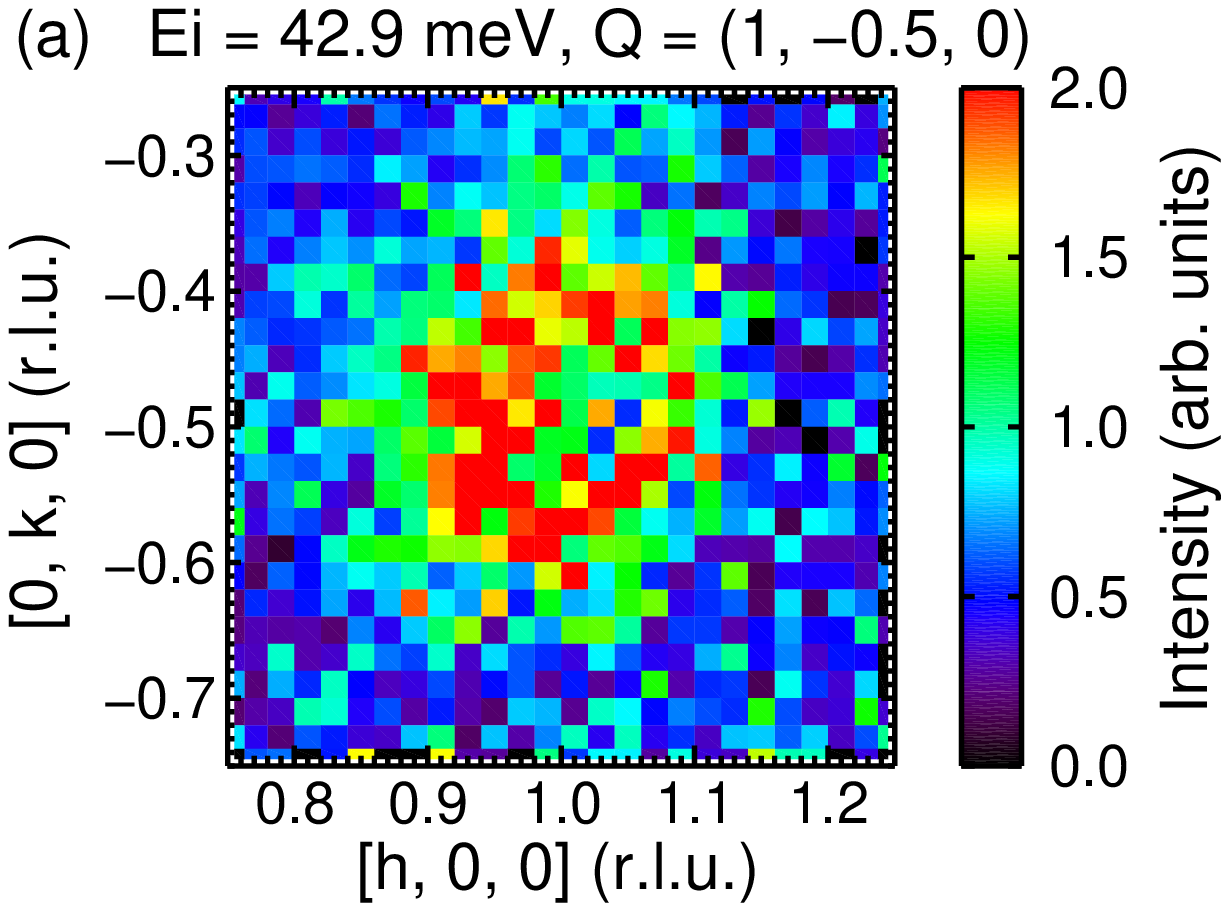}
\hspace{1pc}
\includegraphics[width=0.48\hsize]{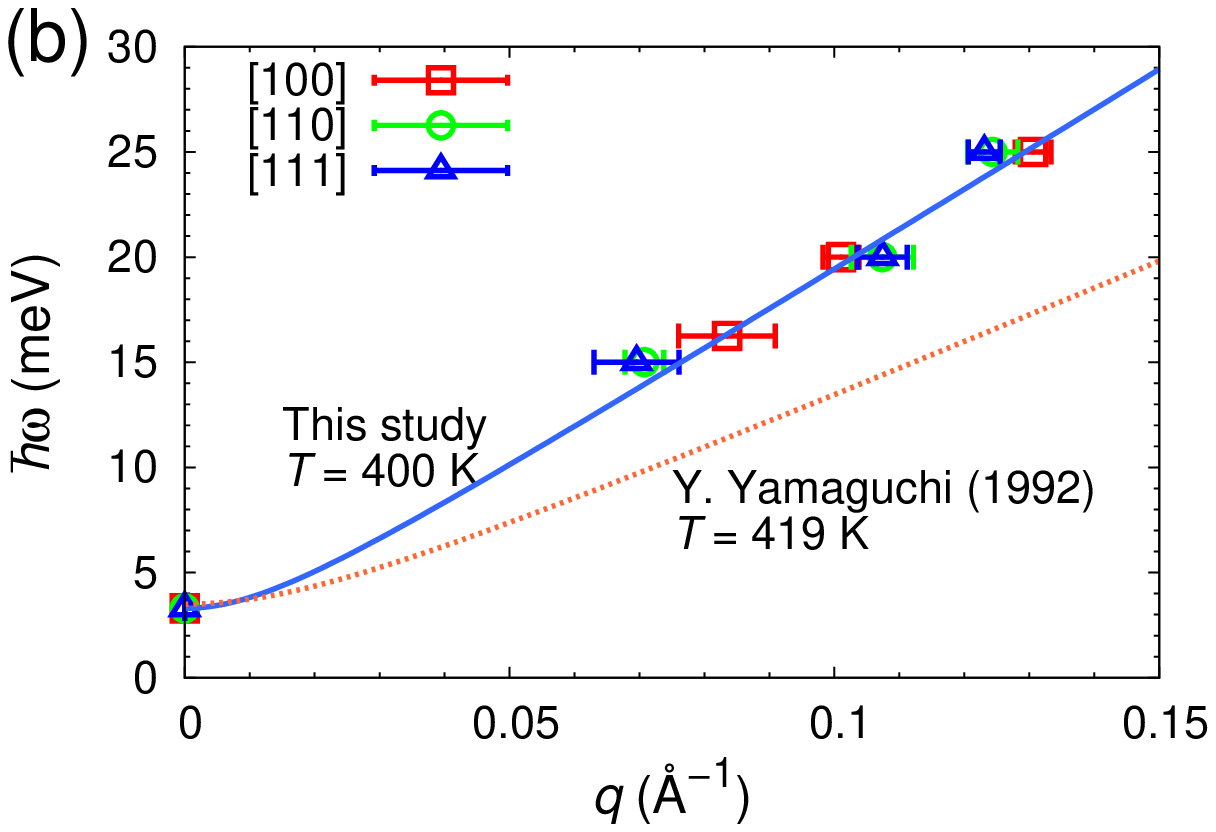}\\
\end{center}
\hspace{0.5pc}
\includegraphics[width=0.45\hsize]{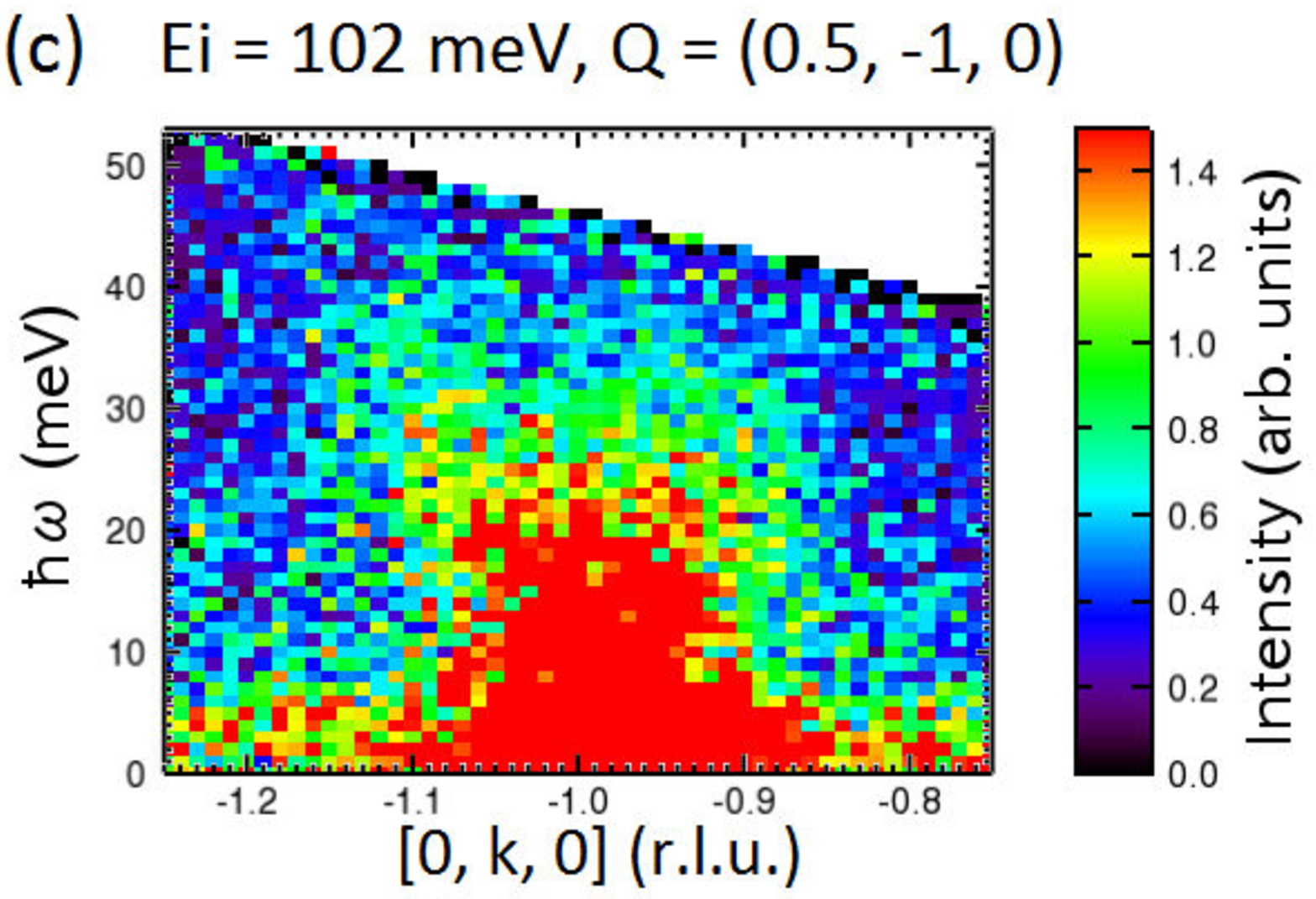}
\caption{\label{fig3}(a) Intensity map with $h$ and $k$ around $\ibbs{Q} = (1, -0.5, 0)$ at ${\hw} = 20$~meV with ${\Ei} = 42.9$~meV. (b) The spin wave dispersion along [100], [110] and [111] direction. The solid line shows the fit with the equation, $({\hw})^2 = c^2q^2 + \Delta^2$. The dotted line denotes the dispersion curve at $T = 419$~K reported in~\cite{YamaguchiY1992}. (c) Intensity map with {\hw} and $k$ around $\ibbs{Q} = (0.5, -1, 0)$ with ${\Ei} = 102$~meV.}
\end{figure}
Second, the spin wave dispersion was investigated by using high-energy incident neutrons. Figure~\ref{fig3}(a) shows the typical intensity map with $h$ and $k$ at ${\hw} = 20$~meV. The spin wave dispersion along [100], [110] and [111] direction was estimated and shown in Fig.~\ref{fig3}(b). The dispersion was isotropic within the measurement accuracy.
A fitting with the equation, $({\hw})^2 = c^2q^2 + \Delta^2$, yields the spin wave stiffness $c = 190(10)$~meV{\AA}, where $q$ is the wave number away from a W point, and $\Delta$ is an energy gap fixed to 3.3~meV. 
At 419~K, $c$ and $\Delta$ were reported to be 130~meV{\AA} and 3.5~meV by the previous report~\cite{YamaguchiY1992}. The result indicates that the decreasing temperature from 419~K = 0.88{\TN} to 400~K = 0.84{\TN} leads to enhance $c$ as much as 50\%, while it has little influence on $\Delta$.
Figure~\ref{fig3}(c) provides the intensity map with {\hw} and $k$ in high-energy region. The dispersion does not reach the zone boundary at $k = -1.25$ and -0.75~r.l.u. even at ${\hw} = 40$~meV. 
It was reported that the dispersion reaches the zone boundary at about 30~meV at 459 K = 0.97{\TN}~\cite{TomiyoshiS1992}. The result indicates that the spin wave energy at the zone boundary increases with decreasing temperature, which is consistent with the increase of $c$ with decreasing temperature.

\section{Discussion}
A large reduction in $c$ with increasing temperature was observed even in the F-phase, which will be an indication of the itinerancy of the magnetic moments. This character is similar to that observed in another 5$d$ alloy, FePt$_3$~\cite{KohgiM1980}.
On the other hand, if we discuss it within the molecular-field approximation, the variation in $c$ with keeping ${\Delta}$ constant indicates a variation in the exchange interaction between the spins. Suppose that ${\Delta}$ originates from single-ion anisotropy, the unchanged ${\Delta}$ indicates that both the anisotropy constant and the moment size remain constant. The previous study of the temperature dependence of the moment size~\cite{YasuiH1992} supports the result; 3.4 and 3.2${\mu}_{\rm B}$ at $T = 400$ and 419~K, respectively.
$c$ depends on moment size and exchange interaction. Therefore, a variation in $c$ result in a variation in the exchange interaction. 
The previous inelastic neutron scattering study~\cite{YamaguchiY1992} found that the exchange interaction is different between the F- and D- phase, and pointed out that the difference originated in the band structure. Our result shows that the exchange interaction varies even in the F-phase, suggesting that the dynamical magnetic properties of metallic {\m} in the F-phase could not be fully understood under the molecular-field approximation with constant exchange interaction, and the itinerant nature of the magnetic electron should be taken into account.
Further study on the temperature dependence of the spin wave will be needed by using the same sample to rule out the capability of the sample dependence.

Another magnetic structure for the high-temperature phase was proposed by Long~\cite{LongMW1991}.
This is a non-collinear magnetic structure like the D-phase. All the Mn moments are ordered along twelve [110] directions.
Neutron diffraction measurements cannot distinguish these two magnetic structures, because they have identical magntic peak positions and intensities.
It was proposed~\cite{HiranoY1995} that three acoustic modes will be exist in this non-collinear structure, and they are all gapped.
On the other hand, the existence of the two acoustic modes, gapped and gapless mode, was expected in the partial disordered state. The gapless mode corresponds to the rotation of all the spins simultaneously by the same amount in the $bc$-plane in Fig.~\ref{fig1}(a).
In our result, the existence of the gapless mode is unclear as we can see in Fig.~\ref{fig2}(a).
Although the tetragonal distortion in the F-phase~\cite{IkedaT2003} supports the partial disordered state, further investigation into the existence of the gapless mode will be useful to establish the magnetic structure.

\section{Summary}
Spin wave dispersion in the F-phase of {\m} was investigated just above the transition temperature by using the neutron scattering technique. The spin wave stiffness at $T = 400$~K is 50\% harder than that at $T = 419$~K~\cite{YamaguchiY1992}, while the spin wave energy gap was unchanged.
The dramatic change in the spin wave stiffness indicates the itinerancy of the magnetic moments in the F-phase.
This study has demonstrated that the magnetic properties of {\m} should be described by the intermediate character between itinerant electron magnets and localized spin magnets.

\ack
One of the author (S. Ibuka) is indebted to Y. Ishikawa and K. Namba at KEK for supporting the chemical analysis. Part of this work was performed using the facilities of CROSS-Tokai user laboratories, and the facilities of the Institute for Solid State Physics, the University of Tokyo. The neutron experiment at MLF in J-PARC was performed under a user program (Proposal No. 2014B0153).

\section*{References}
\bibliography{Mn3Pt}

\providecommand{\newblock}{}
\begin{thebibliography}{10}
\expandafter\ifx\csname url\endcsname\relax
  \def\url#1{{\tt #1}}\fi
\expandafter\ifx\csname urlprefix\endcsname\relax\def\urlprefix{URL }\fi
\providecommand{\eprint}[2][]{\url{#2}}

\bibitem{KamiharaY2008}
Kamihara Y, Watanabe T, Hirano M and Hosono H 2008 {\em J. Am. Chem. Soc.\/}
  {\bf 130} 3296--3297

\bibitem{KohgiM1980}
Kohgi M and Ishikawa Y 1980 {\em J. Phys. Soc. Jpn.\/} {\bf 49} 985--993

\bibitem{IshikawaY1978}
Ishikawa Y 1978 {\em J. Appl. Phys.\/} {\bf 49} 2125--2130

\bibitem{KrenE1967}
Kr\'{e}n E, K\'{a}d\'{a}r G, P\'{a}l L and Szab\'{o} P 1967 {\em J. Appl.
  Phys.\/} {\bf 38} 1265--1266

\bibitem{KrenE1968a}
Kr\'{e}n E, K\'{a}d\'{a}r G, P\'{a}l L, S\'{o}lyom J, Szab\'{o} P and
  Tarn\'{o}czi T 1968 {\em Phys. Rev.\/} {\bf 171} 574--585

\bibitem{IkedaT2003}
Ikeda T and Tsunoda Y 2003 {\em J. Phys. Soc. Jpn.\/} {\bf 72} 2614--2621

\bibitem{YasuiH1992}
Yasui H, Ohashi M, Abe S, Yoshida H, Kaneko T, Yamaguchi Y and Suzuki T 1992
  {\em J. Magn. Magn. Mater.\/} {\bf 104-107} 927--928

\bibitem{RicodeauJA1974}
Ricodeau J~A 1974 {\em J. Phys. F: Metal Phys\/} {\bf 4} 1285--1303

\bibitem{TomiyoshiS1990}
Tomiyoshi S, Yasui H, Kaneko T, Yamaguchi Y, Ikeda H, Todate Y and Tajima K
  1990 {\em J. Magn. Magn. Mater.\/} {\bf 90 \& 91} 203--204

\bibitem{YamaguchiY1992}
Yamaguchi Y, Yasui H, Funahashi S, Yamada M, Ohashi M and Kaneko T 1992 {\em
  Physica B\/} {\bf 180 \& 181} 235--237

\bibitem{TomiyoshiS1992}
Tomiyoshi S, Kaneko T, Steigenberger U, Chapell A~J, Hagen M and Todatede Y
  1992 {\em Physica B\/} {\bf 180 \& 181} 227--229

\bibitem{IkedaT2004}
Ikeda T and Tsunoda Y 2004 {\em J. Magn. Magn. Mater.\/} {\bf 272-276} 482--484

\bibitem{TomiyasuK2012}
Tomiyasu K, Yasui H and Yamaguchi Y 2012 {\em J. Phys. Soc. Jpn.\/} {\bf 81}
  114724--1--4

\bibitem{KajimotoR2011}
Kajimoto R, Nakamura M, Inamura Y, Mizuno F, Nakajima K, Ohira-Kawamura S,
  Yokoo T, Nakatani T, Maruyama R, Soyama K, Shibata K, Suzuya K, Sato S,
  Aizawa K, Arai M, Wakimoto S, Ishikado M, ichi Shamoto S, Fujita M, Hiraka H,
  Ohoyama K, Yamada K and Lee C~H 2011 {\em J. Phys. Soc. Jpn.\/} {\bf 80}
  SB025--1--6

\bibitem{NakamuraM2009}
Nakamura M, Kajimoto R, Inamura Y, Mizuno F, Fujita M, Yokoo T and Arai M 2009
  {\em J. Phys. Soc. Jpn.\/} {\bf 78} 093002--1--4

\bibitem{InamuraY2013}
Inamura Y, Nakatani T, Suzuki J and Otomo T 2013 {\em J. Phys. Soc. Jpn.\/}
  {\bf 82} SA031--1--9

\bibitem{dave}
Azuah R~T, Kneller L~R, Qiu Y, Tregenna-Piggott P~L~W, Brown C~M, Copley J~R~D
  and Dimeo R~M 2009 {\em J. Res. Natl. Inst. Stan. Technol.\/} {\bf 114}
  341--358

\bibitem{LongMW1991}
Long M~W 1991 {\em J. Phys.: Condens. Matter\/} {\bf 3} 7091--7094

\bibitem{HiranoY1995}
Hirano Y, Suzuki N, Shirai M and Motizuki K 1995 {\em J. Magn. Magn. Mater.\/}
  {\bf 140-144} 1975--1976

\end{thebibliography}

\end{document}